\documentclass[runningheads]{llncs}

\makeatletter
\newcommand{\@chapapp}{\relax}%
\makeatother

\usepackage{graphicx}
\usepackage{todonotes,citesort,verbatim}
\usepackage[linesnumbered,ruled,vlined]{algorithm2e}
\usepackage{ulem}
\usepackage{wrapfig,paralist}
\usepackage{appendix}
\usepackage{url}
\usepackage{cite}
\usepackage[pdfpagelabels,colorlinks,allcolors=blue]{hyperref}

\begin{document}
\title{Visualizing The Intermediate Representation of Just-in-Time Compilers}
\titlerunning{Visualizing IR of JIT Compiler}
%
\author{HeuiChan Lim \and
Stephen Kobourov}
\authorrunning{H.~Lim and S.~Kobourov}
%
\institute{
Department of Computer Science, University of Arizona}
\maketitle              

\begin{abstract}
Just-in-time (JIT) compilers are used by many modern programming systems in order to improve performance. Bugs in JIT compilers provide exploitable security vulnerabilities and debugging them is difficult as they are large, complex, and dynamic. Current debugging and visualization tools deal with static code and are not suitable in this domain. 
We describe a new approach for simplifying the large and complex intermediate representation, generated by a JIT compiler and visualize it with a metro map metaphor to aid developers in debugging.

Experiments using our prototype implementation  on Google’s V8 JavaScript interpreter and TurboFan JIT compiler demonstrate that it can help identify and localize buggy code.

\end{abstract}

\section{Introduction}\label{sec:introduction}

Many modern programming systems, such as JavaScript engines that are running our web browsers, use just-in-time (JIT) compilers to improve performance. The most well-known web browsers that we use every day are Google Chrome's V8~\cite{v8}, which has 2.65 billion users~\cite{chromestat}, Microsoft Edge's ChakraCore~\cite{chakracore}, which has 600 million users~\cite{edgestat}, Mozilla Firefox's SpiderMonkey~\cite{spidermonkey}, which has 222 million users~\cite{firefoxstat}, and Apple Safari's WebKit~\cite{javascriptcore}, which has 446 million users\cite{safaristat}, etc., Additionally, programming languages such as Java, C\#, and Ruby, also use JIT compilers to improve performance.

Meanwhile, the JIT compiler bugs can lead to exploitable security vul\-ner\-abilities~\cite{rabet2017,issue5129,issue8056,issue1072172,issue961237}. For example, a JIT compiler bug in Google V8 was found by the Microsoft Offensive Security Research  team (CVE-2017-5121~\cite{rabet2017}). Fortunately, this was reported quickly and fixed by V8 team, but hackers could have used it  to hijack Chrome user data (i.e., passwords), and to navigate to other sites and execute malicious programs. Therefore, it is important to analyze and localize the bug quickly. However, existing work and available tools focus on static code~\cite{DBLP:conf/pldi/HolzleCU92,DBLP:conf/pldi/BrooksHS92,DBLP:conf/pldi/Adl-TabatabaiG96}, so they are not suitable for developers in debugging the JIT compiler, which generates code at run-time. Additionally, the size and complexity of JIT-based systems\cite{hinkelmann2017understanding} combined with the dynamic nature of JIT compiler optimizations, make it challenging to analyze and locate bugs quickly. For example, Google V8 has more than 2,000 source files and more than 1 million lines of code, and Apple's Webkit has approximately 22,000 source files and 500,000 lines of code. 

Traditional debuggers rely on text even though the main feature of a JIT compiler is building a graph-like structure to translate bytecode into optimized machine code. With this in mind, we propose a new debugging aid tool, which visualizes the JIT compiler's intermediate representation (IR) that it generates and optimizes. Our approach uses IR identification and generation techniques described by Lim and Debray~\cite{DBLP:conf/vee/LimD21}, which discusses the compiler-related half of the visualization tool's pipeline. This paper focuses on the visualization half, which includes: (1) merging multiple IR graphs into a single graph; (2) simplifying the merged graph; (3) converting the simplified graph into a hypergraph; (4) simplifying the hypergraph; and (5) visualizing the hypergraph using a metro map metaphor~\cite{DBLP:journals/tvcg/JacobsenWKN21}.

The resulting  visualization of the intermediate representation of the JIT compiler allows developers to answer questions such as:

\begin{compactenum}
\item What optimizations took place to generate the machine code?
\item What is the relationship among the optimization phases?
\item Which optimization phase was most active?
\item With a specific node that represents some operation, which optimization(s) took place (e.g., what phase optimized an arithmetic operator node for a subtraction)?
\item Which optimization phases are likely to be buggy?
\end{compactenum}

\section{Related Work}\label{sec:related-work}

Although there is a considerable body of work on debugging approaches for static code compilers and optimized code, there is very little work on using the intermediate representation and visualizing it to show the explicit information about the compilation and optimization processes.

Google V8's Turbolizer~\cite{lukeolney2019turbolizer,jimenez2020intro} is one of very few IR visualization tools. It shows the final IR graph after each optimization process and provides interactive features to view the control flow graphs for each optimization phase.  Although Turbolizer provides some information about the IR nodes and their relationships, it does not provide enough information about the optimization process and cannot answer several of our initial set of questions.

Dux {\it et al.}~\cite{DBLP:conf/iwpc/DuxIDFK05} introduced an approach to visualize dynamically modifying code at run-time in call graph and control flow graphs. Their tool shows the changes in graph with an animation of the underlying graph, allowing  end-to-end play, pause, and forward/backward step-by-step animation.
Colberg {\it et al.}~\cite{DBLP:conf/softvis/CollbergKNPW03} proposed 
GEVOL: a system that visualizes the evaluation of software. It uses the CVS version control system to get the information about how a program has been changed or evolved over time and visualizes it as a dynamic graph. 
CFGExplorer~\cite{DBLP:journals/cgf/DevkotaI18} visualizes the control flow graph of a program to represent the program structure for dynamic binary analysis. It provides interactive features where users can find specific memory addresses, loops, and functions that they are interested to analyze the system.
CcNav~\cite{DBLP:journals/tvcg/DevkotaAKLI21}  analyzes and visualizes a C++ compiler's optimization process with a call graph, control flow graph, and loop hierarchies.

Control flow graph and call graphs are popular in program analysis, especially, when analyzing  static code, with a 
code-to-compiler-to-assembler pipeline. However, these graphs are different from the intermediate representation generated dynamically from bytecode by JIT compilers. Tools for visualizing and interacting with control flow graph and call graphs (such as those above) are not sufficient for visualizing the IR graph as, e.g., they cannot capture the optimization phases.

\section{Background}\label{sec:background}

Here we briefly introduce several concepts relevant to JIT compilers and to our visualization.

\textbf{Interpreter}: An interpreter is a computer program that converts input source code into bytecode and executes it without compiling it into a machine code~\cite{DBLP:conf/hpcn/GreggEK01}.

\begin{figure}[t]
    \begin{center}
    \includegraphics[width=0.6\textwidth]{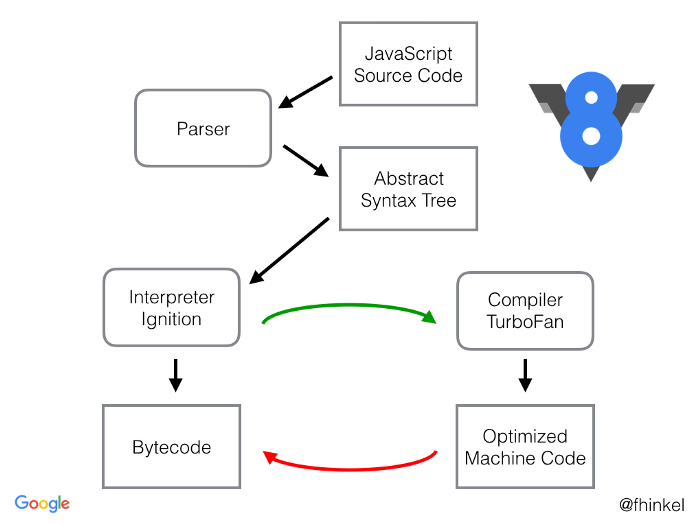}
    \includegraphics[width=0.39\textwidth]{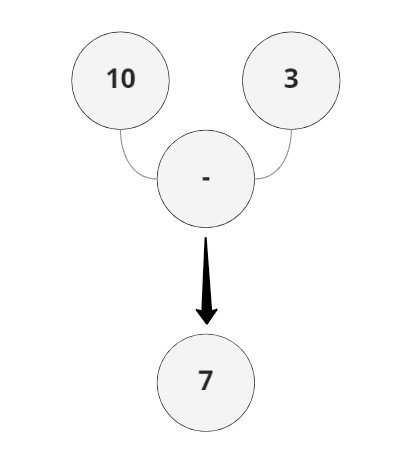}
    \end{center}
    \caption{(a) V8 Pipeline~\cite{hinkelmann2017understanding} (b) Example of constant folding optimization: if the compiler determines that the value of \textit{10-3} will not change during the code execution, it is more efficient to replace the 3-node subgraphs with a single node.}
    \label{fig:v8piepline_optexample}
\end{figure}

\textbf{Just-in-Time (JIT) compiler}: A just-in-time (JIT) compiler is a program that turns bytecode into instructions that can be sent directly to a computer's processor, which is widely used to improve performance~\cite{DBLP:journals/concurrency/IshizakiKYTOSOKN00}. Google V8 is an example of a JavaScript engine that has a JIT compiler in the pipeline; see Fig.~\ref{fig:v8piepline_optexample} (a). Other systems, such as ChakraCore, JavaScriptCore, and Java Runtime Environment, have similar pipelines.

\textbf{Bytecode}: Bytecode is an instruction generated from input source code by an interpreter. Bytecode is portable, unlike compiled programs, so they are widely used in modern languages and systems, such as JavaScript, Python, and Java, etc.\cite{DBLP:conf/jit/Dahm99}.

\textbf{Optimized code}: Optimized code is an instruction generated from bytecode by a JIT compiler. Optimized code is machine code that can be directly executed by a processor. Unlike bytecode, optimized code is system-specific, and much faster in execution.

\textbf{Intermediate Representation (IR)}: The IR is type of representation known as sea-of-nodes~\cite{10.1145/3178372.3179503,meurer2016v8,Sevcik2016keyinstructions}. Unlike other graphs used in the program analysis field, such as control-flow or data-flow graphs which have specific types of nodes, nodes in the sea-of-nodes graph represent different types of nodes: from scalar values and arithmetic operators to variables and control-flow nodes and function entry nodes. Consider, for example, the  expression $\it var\ myVar = 100;$ optimized at phase $\alpha$. This generates 3 different types of nodes: a variable node for $\it myVar$, an arithmetic operator node for {\it Equal}, and a scalar value node for  constant $\it 100$. Other types of nodes in the IR include control-flow nodes (e.g., $\it call,\ branch,\ switch$) and function entry nodes (e.g., $\it Start,\ Return,\ Checkpoint$).

\textbf{Optimization}: In the context of JIT compilers, optimization means adding, removing, and merging nodes and edges in the graph during the execution. In a single JIT compilation, the JIT compiler executes several different phases of optimization (i.e., inlining, loop peeling, constant propagation) to generate efficient machine code; see Fig.~\ref{fig:v8piepline_optexample}(b) for a constant propagation example. Each optimization phase modifies the IR graph and corresponds to a new hyperedge (all the nodes generated or optimized in this phase); we store all of them. 

\textbf{Proof-of-Concept Program}: A ``Proof-of-Concept" (PoC) program is an input program that is used to trigger the buggy behavior in the JIT compiler, i.e., a valid program (without any bugs) which when run can reveal bugs in the JIT compiler. In our experiment, we are targeting JavaScript engine V8, so the PoC that we are using is a JavaScript program.

\textbf{Instruction-level trace}: An instruction-level trace is a file that holds all the instructions that a programming system, such as a JIT compiler, has generated and executed at run-time. The instructions are in a machine-level code with symbol information (i.e., function names). Such traces are used for analyzing the executed system's status, such as performance and behavior, as well as for debugging. 

\section{Visualizing the Intermediate Representation }\label{sec:research}

Our approach to capturing and visualizing the intermediate representation of a JIT compiler, starting from the input JavaScript source code, consists of the following steps:

\begin{enumerate}
\item Start by automatically modifying the original input JavaScript program $P_0$, where $P$ is program, to create a set of $N$ similar input programs $\{P_1,...P_N\}$. The likelihood of identifying the suspicious buggy phase increases as $N$ increases (by default we use $N=20$). The program variants are created by first generating the abstract syntax tree for the original input code and then randomly modifying nodes in the tree, with a set of allowable modifications as described in~\cite{DBLP:conf/vee/LimD21}. Consider, for example, an expression that adds two numbers $\it 2 + -1$. We can modify $2$ or $1$ to any other integer. We can change the operation $+$ to another arithmetic operation. All such allowable modifications must pass semantic and syntax checks.
\item Run each program $P_i$ and collect the instruction-level traces.
\item Analyze the traces to determine whether or not $P_i$ manifested the bug and to identify $P_i$'s IR within the JIT compiler, together with the optimization phases executed while optimizing $P_i$.
\item Select candidate hyperedges, suspected to be buggy, from the information gathered in step 3.
\item Merge all selected candidate hyperedges into the original IR from $P_0$.
\item Simplify the merged IR as a graph by reducing the number of nodes and edges.
\item Convert the simplified graph into a hypergraph by extracting the hyperedges from step 4 and analyzing each node's optimization status. Consider, for example, a node $A$  generated in hyperedge $\alpha$ and optimized in hyperedge $\phi$ and $\gamma$; in this step $A$ (which is in $\alpha$ from the start) is added to $\phi$ and $\gamma$.
\item Simplify the hypergraph by reducing the number of hyperedges and nodes.
\item Visualize the simplified hypergraph with MetroSets~\cite{DBLP:journals/tvcg/JacobsenWKN21}.
\end{enumerate}

The compiler-related  steps 1-4 above are covered in detail  in~\cite{DBLP:conf/vee/LimD21} and details for steps 5-9 are in the next section.
\subsection{Intermediate Representation}\label{sec:ir_generation}

Recall that the intermediate representation (IR) of a JIT compiler is a sea-of-nodes graph that the compiler generates at the beginning of its execution by parsing the bytecode and optimizing it with several optimization phases.

Formally, the IR is a simple, undirected graph $G = (V, E)$, where $V$ represents the nodes optimized by the JIT compiler and $E$ contains pairs of nodes connected by different relationships (e.g., semantic and syntax relationships, such as math expressions). By keeping track of the optimization information for each node we construct the hypergraph $H = (V, S)$ from $G$, where $V$ is a set of nodes optimized by the JIT compiler and each hyperedge in $S$ represents an optimization phase. 

Two important node features are phases and opcodes. Phases are the optimization phases where a node was generated and optimized (and which later correspond to hyperedges). Opcodes represent node operations (e.g., add, sub, return). A node also has two different attribute groups: (1) \textit{basic}, such as a node id, address, list of neighbors, opcode, and IR ID (the ID of IR that the node belongs to); and (2) \textit{optimization}, such as hyperedge (phase) ID, hyperedge name, replace/kill/remove/append phase.

\begin{figure}[t]
    \begin{center}
    \vspace{-.5cm}
    \includegraphics[width=0.49\linewidth]{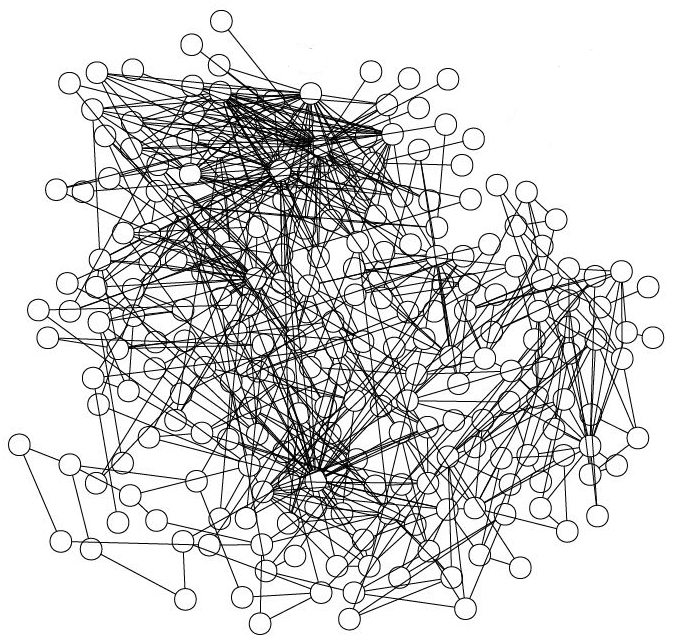}
    \includegraphics[width=0.49\linewidth]{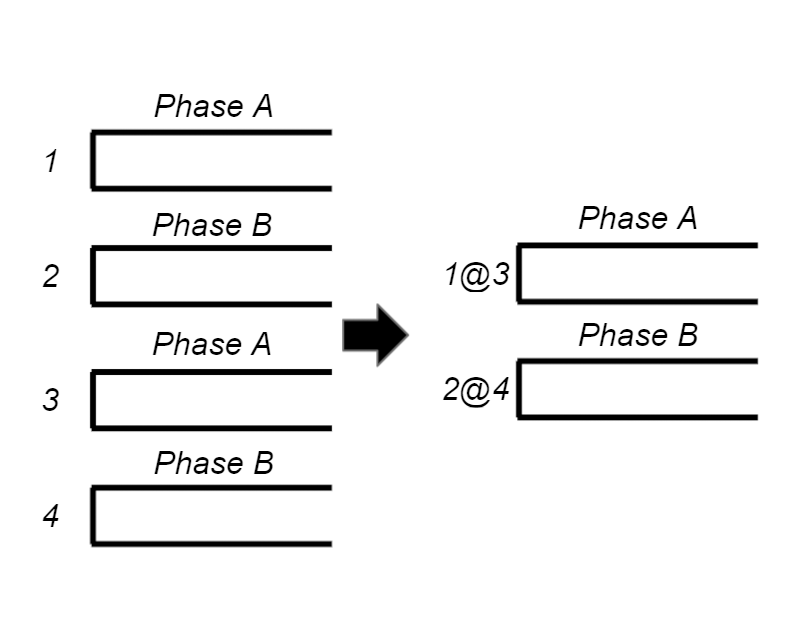}
    \vspace{-.8cm}
    \end{center}
    \caption{(a) Example of an IR graph; (b) example of hypergraph simplification.}
    \label{fig:irgraph_set_merge}
\end{figure}

Recall that given one JavaScript code we generate $N$ similar versions to see if any of them trigger bugs. We generate the IRs for all of these versions (typically about 20). In the real-world examples we work with, each such IR graph has about 300-500 nodes and 30-40 optimization phase executions.
\subsection{Merging Intermediate Representation Hyperedges}\label{sec:graph_mege}

Recall that we now have $N$ intermediate representations from our original JavaScript program. They are similar, but different, and in this step of our pipeline we merge them all into one single graph by merging hyperedges. There are two main reasons for this IR merging, First, we would like to visualize the differences among the graphs in one single view. Second, as shown in~\cite{DBLP:conf/vee/LimD21}, by comparing hyperedges from a buggy program IR to hyperedges from a non-buggy program IR, we can find differences in some hyperedges due to different optimizations, and thus find the bug. Consider, for example, a hyperedge $\alpha$ in both buggy and non-buggy program IRs and suppose that an additional node (the result of incorrect optimization) makes a buggy program's $\alpha$  different from the non-buggy program's $\alpha$.  
A merged hyperedge will show this additional node and its attributes will identify the buggy IR. 
A developer can now see that there was an optimization difference in $\alpha$ between the buggy and non-buggy programs, check whether this difference causes the buggy behavior.

We use the following procedure to merge IRs. Let $R_0$ be the IR from the original program and $\{R'_1,...,R'_N\}$ the IRs from the modified programs. Let $\{r'_1,...,r'_n\}$ be sub-IRs, where $r'_i$ is a subgraph of $R'_i$ when $R'_i\ne R_0$; i.e., $n$ is the number of IRs different from $R_0$ ($n \le N$). Each $r'_i$ holds candidate hyperedges: $R'_i$ hyperedges different from $R_0$'s hyperedges.  
We traverse all sub-IRs, comparing each to  $R_0$, and update the merged IR. 
If $s'_i \in r'_i$ and $s'_i \notin R_0$, where $s'_i$ is a hyperdege, we add $s'_i$ to $R_0$. If $s'_i \in r'_i$ and $s_i \in R_0$, where $s_i$ has same hyperedge name and execution order with $s'_i$, but $s'_i \ne s_i$, we merge $s'_i$ with $s_i$; see Algorithm~\ref{algo:merging} in Supplementary Materials for details.
\subsection{Intermediate Representation Simplification}\label{sec:graph_simplify}

Although a graph with all the nodes and all the edges may be useful for debugging, the complexity, as we can see in Fig.~\ref{fig:irgraph_set_merge} (a) makes it difficult for developers to use. Therefore, we simplify the graph (hopefully without losing much information), convert it into a hypergraph, and simplify the hypergraph (again, hopefully without loosing much information). The main goal is to end up with an interactive visualization that allows developers to debug. Our IR simplification has two parts: (1) Reduce IR as a graph, (2) Reduce IR as a hypergraph.

\subsubsection{Reducing the IR Graph }\label{sec:reducegraph} 
\begin{figure}
    \begin{center}
    \includegraphics[width=\textwidth]{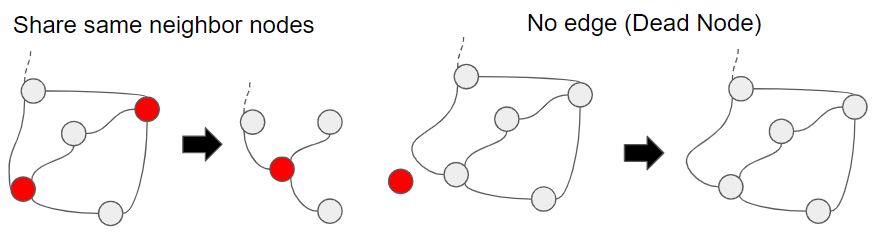}
    \end{center}
    \caption{Reducing the IR Graph: two red nodes that share the same neighbors are merged (left) and a red dead node is removed (right). }
    \label{fig:node_reduce}
\end{figure}

We remove dead nodes (nodes with no adjacent edges) as they are not translated into machine code and do not affect other nodes. We then identify nodes that can be merged (reducing the total number of nodes and edge) without losing important information. A pair of nodes is merged if they have the same opcode, the same optimization information, belong to the same IR (which can be identified by the IR id attribute), and share the same neighbors; see Fig.~\ref{fig:node_reduce}.
We first traverse each node in the graph and check for dead nodes and removes them. We then re-traverse the graph and compare each node to rest of the nodes, looking for pairs that can be merged. Algorithm \ref{algo:simplificationA} shows a pseudo-code for how the graph simplification is implemented.

\subsubsection{Reducing the IR Hypergraph}\label{sec:reducehypergraph} We first convert the simplified graph $G=(V,E)$ into a hypergraph $H=(V,S)$, by extracting hyperedges based on the optimization phases. Recall that a node $A$  generated in phase/hyperedge $\alpha$ and optimized in phase/hyperedge $\phi$ and $\gamma$ now belongs to all three hyperedges. We reduce the hypergraph by merging suitable pairs of hyperedges. Different nodes can have the same hyperedge names as attribute, but different hyperedge id, as this id is assigned based on the execution order. Therefore, we merge hyperedges with the same name into a single-hyperedge while assigning a new unique identifier generated from the original IDs. We use ID concatenation to obtain unique identifiers. Fig.~\ref{fig:irgraph_set_merge}(b) illustrates hyperedge merging. Consider two hyperedges \textit{A} and \textit{B}, which were executed twice in the order shown. The order is used to create the unique IDs. We merge these 4 hyperedges into two larger hyperedges and assign new IDs that were generated by concatenating two IDs delimited with a special character '@'. Algorithm~\ref{algo:alg_set_reduce} provides an overview of this procedure.
 
This method significantly reduces the number of hyperedges that we have to deal with but increases the number of nodes in each hyperedge. Next, we traverse each hyperedge $s$ in $S$, and we use the nodes opcodes to see if they can be merged. Algorithm \ref{algo:simplificationB} shows the hypergraph simplification.
\section{MetroSets}\label{sec:metroset}

MetroSets~\cite{DBLP:journals/tvcg/JacobsenWKN21} is a tool for visualizing hyepergraphs using the  metro map metaphor. It provides an clean way to show the relationships among the hyperedges, which in our case represent the relationships among the optimizations. Further, MetroSets provides simple and intuitive interactions that make is possible to quickly identify hyperedges that contain a suspicious (buggy) node, or hyperedges that intersect with a particular suspicious (buggy) hyperedge. MetroSets is designed for small-to-medium set systems with no more than 500 nodes and no more than 30 hyperedges. Additional internal requirements limit the number of nodes in each hyperedge to at least 2 and require that all hyperedges must intersect at least 1 other hyperedge. Since some of our hyperedges can be singletons, we add a dummy node when needed (which is later ignored). If one or more of our hyperedges do not intersect any others, they can be treated separately and shown next to the main hypergraph.

Each node in the MetroSet map is labeled with its unique id (representing the timeline of node generation). The attributes shown when hovering over a node are phase, opcode, address, graph id, and phase id. A phase attribute tells the user where the node was generated, and it is useful for the node belongs to multiple sets. The user can distinguish the phase that it was generated and the phases where it was optimized. The phase id helps us analyze the order in which nodes along a given line were generated.

\section{Evaluation}\label{sec:evaluation}

The following experiments were done with our prototype system using a machine with 32 cores (@ 3.30 GHz) and 1TB of RAM, running Ubuntu 20.04.1 LTS. We used a dynamic analysis tool built on top of Intel's Pin software~\cite{DBLP:conf/pldi/LukCMPKLWRH05} for program instrumentation and collecting instruction-level execution traces, XED~\cite{xed} for instruction decoding~\cite{xed}. Additionally, we used esprima-python~\cite{esprima} to generate the syntax-tree for JavaScript code, and escodegen~\cite{escodegen} to regenerate the JavaScript code from the syntax-tree. 

Our prototype targets Google's JavaScript engine V8, focusing in particular on TurboFan, V8's JIT compiler.

We use data from the Chromium bug report site; details about the data, bug descriptions, etc. can be found here~\cite{DBLP:conf/vee/LimD21}.
We are able to localize the bugs in all of the listed bug reports.

\begin{table}[h]
\caption{Graph simplification result table}\label{tbl:simplification-result}
\begin{center}
\begin{tabular}{|r|c|c|c|c|}\hline
  {\bf Report\#} 
      & {\bf Original Set\#} 
      & {\bf Reduced Set\#}
      & {\bf Original Element\#}
      & {\bf Reduced Element\#}\\ \hline\hline
    5129 & 38 & 13 (-59.52\%) & 1695 & 355 (-79.06\%)\\ \hline
    8056 & 36 & 13 (-57.50\%) & 371 & 176 (-52.56\%)\\ \hline
    791245 & 31 & 13 (-51.43\%) & 433 & 198 (-54.27\%)\\ \hline
    961237 & 36 & 13 (-57.50\%) & 478 & 191 (-60.04\%)\\ \hline
    1072171 & 40 & 18 (-50.00\%) & 701 & 326 (-53.50\%)\\ \hline
    880207 & 33 & 13 (-54.05\%) & 555 & 235 (-57.66\%)\\ \hline
\end{tabular}
\end{center}
\end{table}

Table \ref{tbl:simplification-result} shows the result of the simplification step on the bug report data. 

On average 50-60\% reductions in the number of nodes and hyperedges is achieved, resulting in manageable inputs sizes for MetroSets.

\begin{figure*}[ht]
    \centering
    \includegraphics[width=\textwidth]{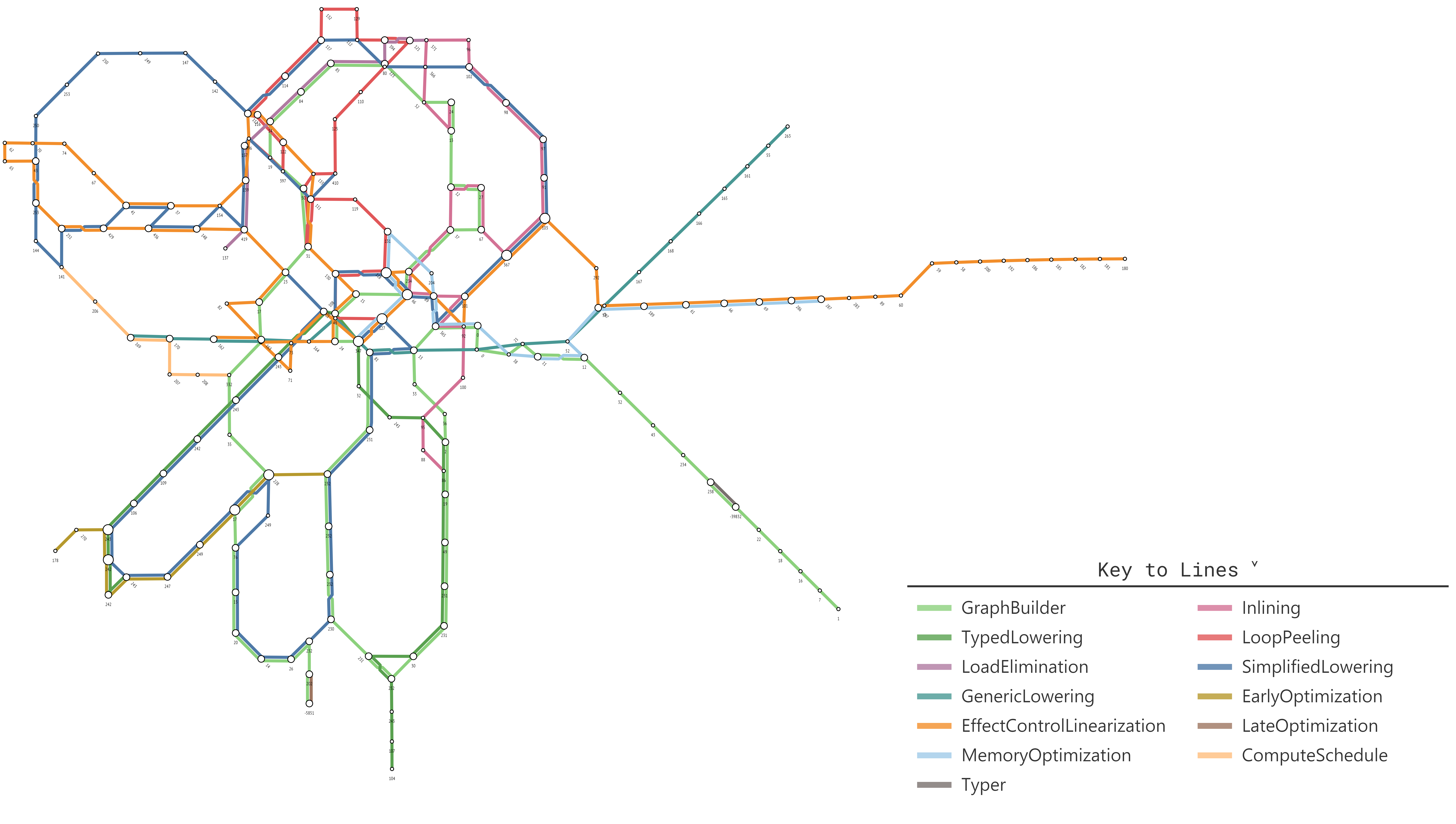}
    \caption{Metro map of IR graph from bug report 5129}
    \label{fig:5129_metro}
\end{figure*}

\bigskip\noindent{\bf Case Study:} Here we briefly go over one example showing how our visualization can assist developers in debugging.

The data comes from JavaScript source code (chromium bug report number 5129). This version of V8's JIT compiler has a problem in the $\it EarlyOptimization$ phase.
We generate 18 additional modified JavaScript codes from the original source code and ran all 19 programs with V8's local executable program called $\it d8$. The resulting  individual instruction trace files are used to generate the IR graph shown in Fig.~\ref{fig:irgraph_set_merge}(a).
Now, we apply our 7-step pipeline to obtain the final metro map representation which shows 335 nodes and 18 sets; see Fig.~\ref{fig:5129_metro}.

We can now attempt to answer some of the questions from Section \ref{sec:introduction}.

\textit{``What optimizations took place to generate the machine code?”} By looking at the set names in ``Key to Lines" legend, we can  identify which optimization phases took place.

\textit{``What’s the relationships among the optimization phases?”} and \textit{``With a specific node that represents some operation, which optimizations took place?"} We can answer these two questions by looking at the lines sharing the nodes. Hovering over the nodes of interest highlights the names of phases in the ``Key to Lines" legend. We can also select the interactive  intersection/union/etc. (exploration modes) to see the relationships among the optimization phases. Fig.~\ref{fig:intersection} shows an example of the relationships between TypedLowering, SimplifiedLowering, and EarlyOptimization, with 3 nodes in their intersection. A developer can investigate the relationship by examining the attributes of each node. The generation order and the hyperedge responsible for the generation can help identify which hyperedge impacted others in the optimization.

\textit{``Which optimization phase was most active?”} We can answer this question by hovering over each line (which reports the number of nodes in the line). The graph builder phase always has the largest number of nodes but it is not an optimization phase. Other lines with many nodes are the most active optimization phases. Fig.~\ref{fig:mostactivehyperedge} shows an example of a most active optimization phase.

\textit{``With a specific node that represents some operation, which optimization(s) took place?"} We can answer this question by hovering over the node of interest. The lines that don't contain the node are grayed out and the displayed node attributes include the opcode, which represents the operation of a node.

\textit{``Which optimization phase is likely to be buggy?”} One natural way to do this is to find parts that differ in the IR graphs with the bug and those without~\cite{DBLP:conf/vee/LimD21}. In other words, a program is  buggy  because either it has additional optimizations or missing optimizations, and this information is captured in the IRs. For example, let's say there is program $A$, which is buggy and program $B$, which is non-buggy. And, both programs were optimized at optimization phase $\gamma$. However, the JIT compiler falsely evaluated the code in program $A$ and omitted some optimization resulting in missing nodes to be translated into machine codes. These nodes are visible in the non-buggy program $B$'s IR as the JIT compiler has correctly performed optimization. Therefore, any line in the metro map that has a high density of nodes from non-original IRs represents that there was a significant difference between the IRs from the buggy and non-buggy programs. In this case study, we have found that the majority of nodes (9 out of 11) in the EarlyOptimization phase line are from different IRs, which are all from non-buggy IRs. This indicates that there was a significant difference in optimization between buggy and non-buggy programs, so the developers are recommended to begin debugging from the EarlyOptimization phase. Additionally, investigating individual nodes in the suspicious line can be helpful. Figure~\ref{fig:nodehover} shows an example of visualized node details that the user can use to learn about the node, such as which bytecode the node corresponds to and  which optimization took place, etc.

\section{Discussion, Limitations, and Future Work}\label{sec:future-work}

We described a new approach for visualizing the intermediate representation of just-in-time compilers using the metro map metaphor and showed how it could be used for debugging. The visualization approach described here is a functional early prototype (available on \url{https://github.com/hlim1/JITCompilerIRViz}) that provides some useful functionality for debugging JIT compilers. There are many limitations and missing features. 

To start with, we target a specific JIT compiler. Generalizing this approach to other JIT compilers (e.g., ChakraCore, SpiderMonkey) can be done by modifying the IR generation process to handle different IR graphs.

After a series of simplifications (of the IR graphs, the merged graph, and the hypergraph) some useful information might be lost. A two-level visualization which shows the simplified hypergraph as an overview but also provides all details on demand will likely be useful.

Currently, we need to hover over each line to identify suspicious phases. This can be improved by applying techniques to localize suspicious phases as described in~\cite{DBLP:conf/vee/LimD21} and automatically highlight/bold such lines.

Along those lines, we might want to provide more detailed information about each node: ``why is this node connected to another?", "what optimization (i.e., removed, added) created this node?", etc. The answers to several such  questions can be collected and provided via appropriate interactions.

\section{Conclusion}\label{sec:conclusion}
In this paper, we have introduced a new implementation for visualizing the intermediate representation of just-in-time compiler generates and optimizes at run-time in a metro map. We have introduced algorithms for graph merging and simplification to effectively reduce the size of a large and complex graph into simpler sets while maintaining the important information to aid the developers in debugging.

\raggedright

\clearpage

\begin{appendices}
\section*{Supplementary Materials}
In this section, we  provide the pseudocode for several of the non-trivial steps in our pipeline, as well several additional examples of buggy code visualized with our system.

\section{Pseudocode}
\IncMargin{1em}
\begin{algorithm}  
 \DontPrintSemicolon
 \SetKwInput{KwData}{Input}
 \KwData{Original IR $R_0 = (V,E)$, Sub-IRs $r' = \{r'_1,...r'_n\}$}
 \KwResult{Merged IR $R^*$}
 \BlankLine
    \SetKwFunction{FMain}{{\it merge\_nodes\_to\_original}}
    \SetKwProg{Fn}{function}{:}{}
    \Fn{\FMain{$R^*,\ r'_i$}}{
        \For {$j = 0$ \KwTo ${\it size}(r'_i)-1$} {
            \If {$v'_j.hyperedge$ exists in $R^*.hyperedges$} {
                add $v'_j$ to $R^*.hyperedges\_to\_nodes$\;
            }
            \Else{
                add $v'_j.hyperedge$ to $R^*.hyperedges$\;
                add $v'_j$ to $R^*.hyperedges\_to\_nodes$
            }
        }
        \textbf{return} $R^*$
    }
 \BlankLine
 \Begin {
    $R^* = R_0$\;
    \For{$i = 0$ \KwTo ${\it size}(r')-1$} {
        $R^* = {\it merge\_nodes\_to\_original(R^*, r'_i)}$\;
    }
}
\caption{Intermediate Representation merging}
\label{algo:merging}
\end{algorithm}
Algorithm~\ref{algo:merging} shows how to merge several sub-IRs (each of which contains candidate hyperedges) into the IR of the original JavaScript program. While traversing the sub-IRs, we call a  $merge\_nodes\_to\_original$ function to update the copy ($R^*$) of original IR $R_0$. The function updates $hyperedges$, which holds the list of hyperedge names, and $hyperedges\_to\_nodes$, which is a map between hyperedges and the list of nodes.

\clearpage

\IncMargin{1em}
\begin{algorithm}  
 \DontPrintSemicolon
 \SetKwInput{KwData}{Input}
 \KwData{Merged IR $R^* = (V^*,E^*)$}
 \KwResult{Simplified IR $R^{**}$}
 \BlankLine
    \SetKwFunction{FMain}{{\it Remove\_Dead\_Nodes}}
    \SetKwProg{Fn}{function}{:}{}
    \Fn{\FMain{$R^*$}}{
        \For {$i = 0$ \KwTo ${\it size}(V^*)-1$} {
            \If {$v^*_i$ has no edge or has only self looping edge} {
                remove $v^*_i$ from $R^*$
            }
        }
        \textbf{return} $R^*$
    }
    \SetKwFunction{FMain}{{\it Remove\_Node\_Edges}}
    \SetKwProg{Fn}{function}{:}{}
    \Fn{\FMain{$R^*$}}{
        $R^{**}$ = copy of $R^*$\;
        $removed\_nodes = \emptyset$\;
        \For {$i = 0$ \KwTo ${\it size}(V^*)-1$} {
            \If{$v^*_i$ not in $removed\_nodes$} {
                \For {$j = i+1$ \KwTo ${\it size}(V^*)-1$} {
                    \If{$v^*_j$ not in $removed\_nodes$ and $v^*_i.properties\ equal\ v^*_j.properties$} {
                        add $v^*_j$ to $removed\_nodes$\;
                        remove $v^*_j$ from $R^{**}$
                    }
                }
            }
        }
        \For {$i = 0$ \KwTo ${\it size}(V^{**})-1$} {
            \For {$j = 0$ \KwTo ${\it size}(v^{**}.edges)-1$} {
                \If{$v^{**}.edges_i$ in $removed\_nodes$}{
                    remove $v^{**}.edges_i$ from $v^*.edges$\;
                }
            }
        }
        \textbf{return} $R^{**}$
    }
 \BlankLine
 \Begin {
    $R^* = Remove\_Dead\_Nodes(R^*)$\;
    $R^{**} = Remove\_Node\_Edges(R^*)$\;
 }
\caption{IR Simplification as Graph}
\label{algo:simplificationA}
\end{algorithm}
Algorithm~\ref{algo:simplificationA} shows how we simplify the merged IR as a graph, by removing nodes and edges.  $Remove\_Dead\_Nodes$ seeks nodes with no adjacent edges (or only with self-loop edges) and removes them from the IR. Then,  $Remove\_Node\_Edge$ merges pairs of nodes if they have the same opcode, the same optimization information, belong to the same IR (which can be identified by the IR id attribute), and share the same neighbors.

\clearpage

\IncMargin{1em}
\begin{algorithm}  
 \DontPrintSemicolon
 \SetKwInput{KwData}{Input}
 \KwData{Simplified graph $R^{**} = (V^{**},E^{**})$}
 \KwResult{Hypergraph $H=(V, S)$}
 \BlankLine
    \SetKwFunction{FMain}{{\it Construct\_HyperGraph}}
    \SetKwProg{Fn}{function}{:}{}
    \Fn{\FMain{$R^{**}$, $R^{**}.hyperedges$}}{
        $V = \emptyset$\;
        $S = \{s_1, s_2, ..., s_N\}$ // Where N is a total number hyperedges in $R^{**}.hyperedges$ and each $s$ represents hyperedges.\;
        $H\ =\ (V,\ S)$\;
        \For {$i = 0$ \KwTo ${\it size}(V^{**})-1$} {
            \If {$v^*_i.hyperedge$ in $R^{**}.hyperedges$} {
                $v = v^*_i$ // $v$ stands for element.\;
                identify the appropriate set $s$ in $S$ and add $v$ to $s$\;
                add $v$ to $V$\;
            }
            \While {$v^*_i.optimized\_hyperedges$} {
                $w = v^*_i$\;
                identify the appropriate set $s$ in $S$ and add $v$ to $s$\;
                add $v$ to $V$\;
            }
        }
        \textbf{return} $H$
    }
 \BlankLine
 \Begin {
    $H = Construct\_HyperGraph(R^{**}, G^{**}.phases)$\;
 }
\caption{Hypergraph Construction}
\label{algo:hypergraphconstruction}
\end{algorithm}
Algorithm~\ref{algo:hypergraphconstruction} shows how we construct a hypergraph $H=(V,S)$ from the simplified graph $G=(V,E)$. First, we extract hyperedges from the simplified graph, which group nodes based on their generation, and add the nodes to $V$. Then we add nodes to all their corresponding hyperedges.

\clearpage
\IncMargin{1em}
\begin{algorithm}  
 \DontPrintSemicolon
 \SetKwInput{KwData}{Input}
 \SetKwFunction{newphase}{newphase}
 \KwData{Hypergraph $H=(V,S)$}
 \KwResult{Hyperedge reduced IR $H^*=(V^*,S^*)$}
 \BlankLine
 \Begin {
    $V^* = V$\;
    $S^* = \emptyset$\;
    $H^* = (V^*,S^*)$\;
    $merged\_hyperedges = \emptyset$\;
    \For{$i = 0$ \KwTo ${\it size}(S)-1$} {
        \If {$s_i$ is not in $merged\_hyperedges$} {
            $concatenated\_id = s_i.id$\;
            $merged\_node\_list[s_i.id] = s_i.nodes$\;
            add $s_i.name$ to $merged\_hyperedges$\;
            \For {$j = i+1$ \KwTo ${\it size}(S)-1$} {
                \If{$s_i.name == s_j.name$}{
                    $j_id = s_j.id$\;
                    add $j_id$ with $@$ symbol to $concatenated\_id$\;
                    add $s_j.nodes$ to $merged\_node\_list[s_i.id]$\;
                }
            }
            add $merged\_node\_list[s_i.id]$ to $S^*[concatenated\_id]$\;
        }
    }
}
\caption{Reducing the Hypergraph}
\label{algo:alg_set_reduce}
\end{algorithm}
Algorithm~\ref{algo:alg_set_reduce} shows the steps for reducing the hypergraph $H=(V,S)$. This is accomplished by merging hyperedges with the same name. Specifically, when merging hyperedges, we generate a new ID by concatenating the IDs delimited with a special character `@' for the merged hyperedge. The corresponding nodes in the hyperedges also get merged.
\clearpage

\IncMargin{1em}
\begin{algorithm}  
 \DontPrintSemicolon
 \SetKwInput{KwData}{Input}
 \KwData{Hypergraph $H^*=(V^*,S^*)$}
 \KwResult{Simplified hypergraph $H^{**}=(V^{**}, S^{**})$}
 \BlankLine
    \SetKwFunction{FMain}{{\it Simplify\_Hyperedge}}
    \SetKwProg{Fn}{function}{:}{}
    \Fn{\FMain{$s^*$}}{
        $s^{**}$ = copy of $s^*$\;
        $simplified\_nodes = \emptyset$\;
        \For {$i = 0$ \KwTo ${\it size}(s^*.nodes)-1$} {
            \If{$node_i$ not in $simplified\_nodes$} {
                \For{$j = i+1$ \KwTo ${\it size}(s^*.nodes)-1$} {
                    \If{$node_j$ not in $simplified\_nodes$} {
                        \If{$node_i.properties$ equal $node_j.properties$ and $node_i.hyperedges$ equal $node_j.hyperedges$} {
                            add $node_j$ to $simplified\_nodes$\;
                            remove $node_j$ from $s^{**}$
                        }
                    }
                }
            }
        }
        \textbf{return} $s^{**}$
    }
 \BlankLine
 \Begin {
    $V^{**} = \emptyset$\;
    $S^{**} = \emptyset$\;
    $H^{**}=(V^{**}, S^{**})$\;
    \For{$i = 0$ \KwTo ${\it size}(S^*)-1$}{
        $s^{**} = Simplify\_Hyperedge(s^*_i)$\;
        add all elements in $s^{**}$ to $V^{**}$\;
        add $s^{**}$ to $S^{**}$
    }
 }
\caption{Hypergraph Simplification}
\label{algo:simplificationB}
\end{algorithm}
Algorithm~\ref{algo:simplificationB} shows the steps for node reduction in the hypergraph. While traversing all nodes in the hyperedge-simplified $H^*$, we compare each node to the other nodes that belong to the same hyperedges. We compare the properties (i.e., opcode) of the nodes, and we merge any nodes that are found to be the same. 

\clearpage

\section{More Results and Examples}\label{sec:moreres}

\begin{figure}[h]
    \includegraphics[width=\textwidth]{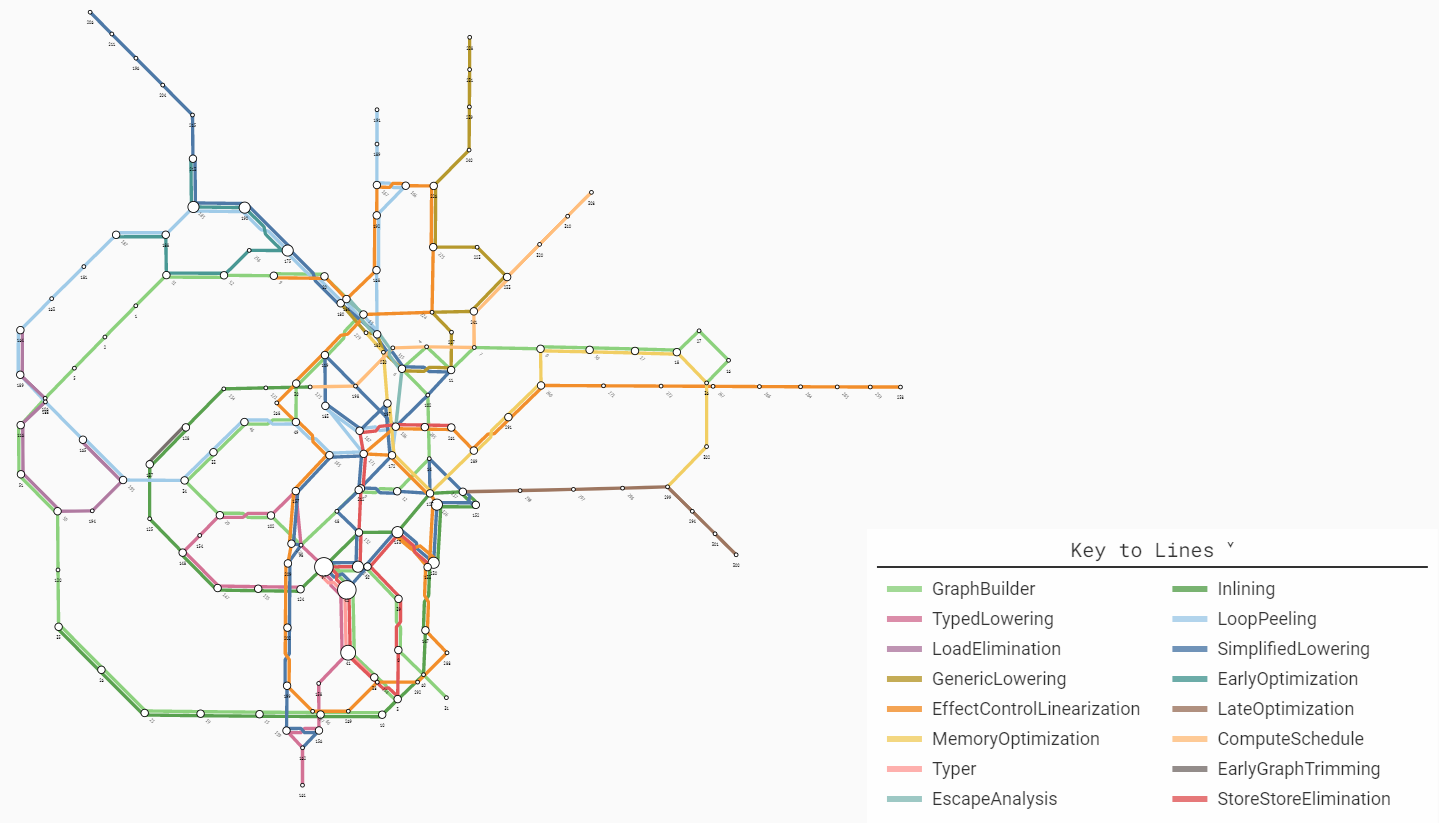}
    \caption{Bug Report 1072171}
    \label{fig:1072171}
\end{figure}

The bug report 1072171 (Figure~\ref{fig:1072171}) is an example with a bug in the Typer hyperedge, where JavaScript code "Math.max" and "Math.min" generates a wrong type due to incorrectly removing one of the typing properties during  optimization~\cite{DBLP:conf/vee/LimD21,issue1072172}.

When we compare the buggy and non-buggy IRs, we can see that the buggy IR is missing a node generation at the Typer hyperedge. 
In other words, when we compare the node generation hyperedges, the Typer hyperedge only exists in the IRs from non-buggy programs, but not in the buggy program IR.
Therefore, we find the node that was was added to the line because of the generation status (and not the optimization status). Specifically, there is only one node, among the four in the Typer line, that was generated by Typer. The other nodes were generated in different phases/lines. Checking the IR ID attribute for this particular node shows that it is from another IR that is non-buggy and all the other nodes on the line are generated from the buggy programs.

\begin{figure}[h]
    \includegraphics[width=\textwidth]{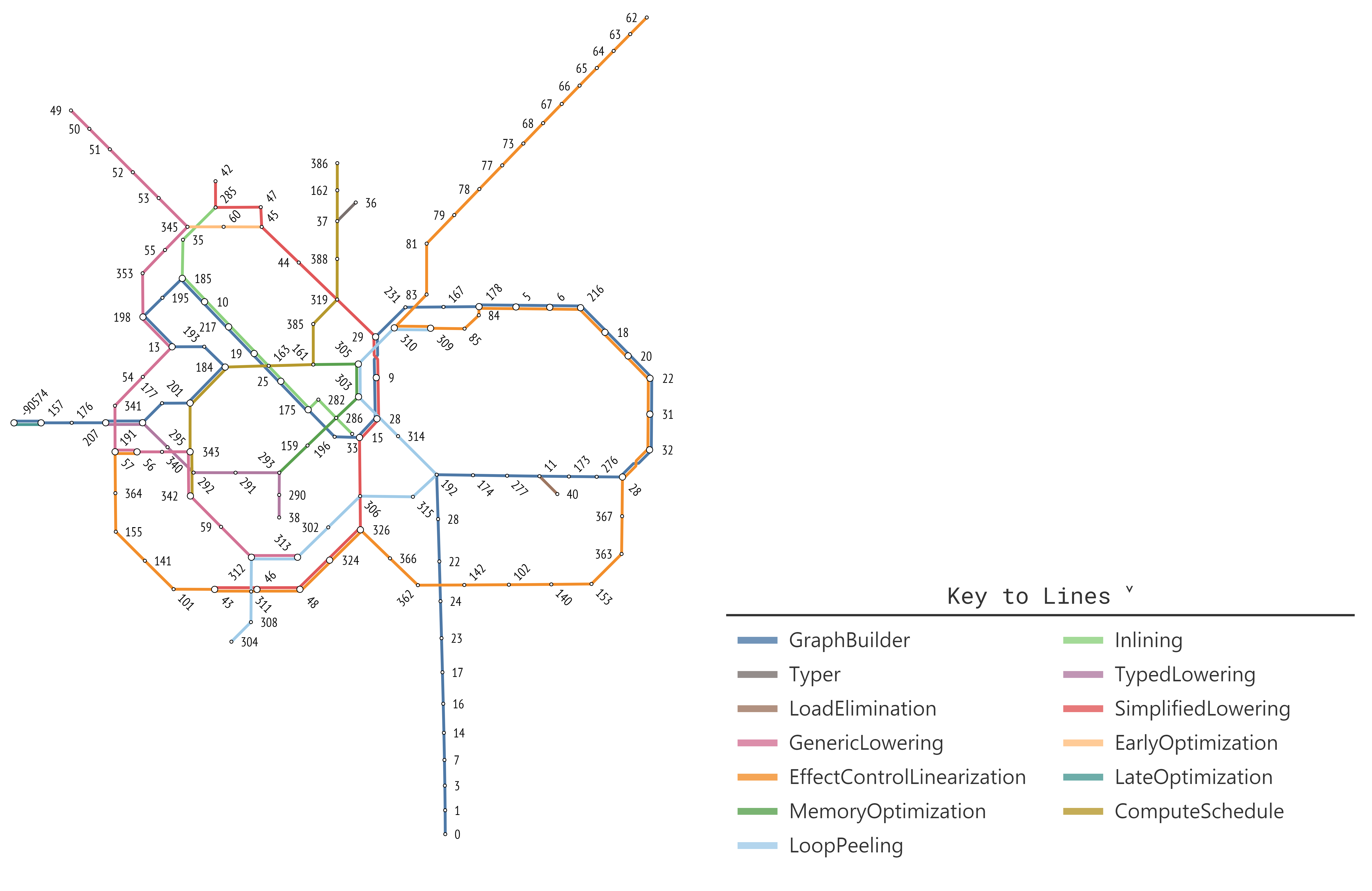}
    \caption{Bug Report 961237}
    \label{fig:961237}
\end{figure}

The bug report 961237 (Figure~\ref{fig:961237}) is an example with a bug in the SimplifiedLowering hyperedge where $null$ is truncated to +0 even in contexts such as $-0\ ==\ null$, due to missing type checking~\cite{DBLP:conf/vee/LimD21,issue961237}.

If a type checking is missing during the evaluation for the optimization, it skips the optimization. Thus, we see a line in which the majority of the nodes belong to other IRs. In this case, the SimplifiedLowering line has 16 nodes, but the majority of the nodes (13 nodes) are from other IRs. This indicates that there was a significant difference between the IR from the original program, which is buggy, and the other IRs from non-buggy programs at SimplifiedLowering hyperedge.

\begin{figure}[h]
    \includegraphics[width=\textwidth]{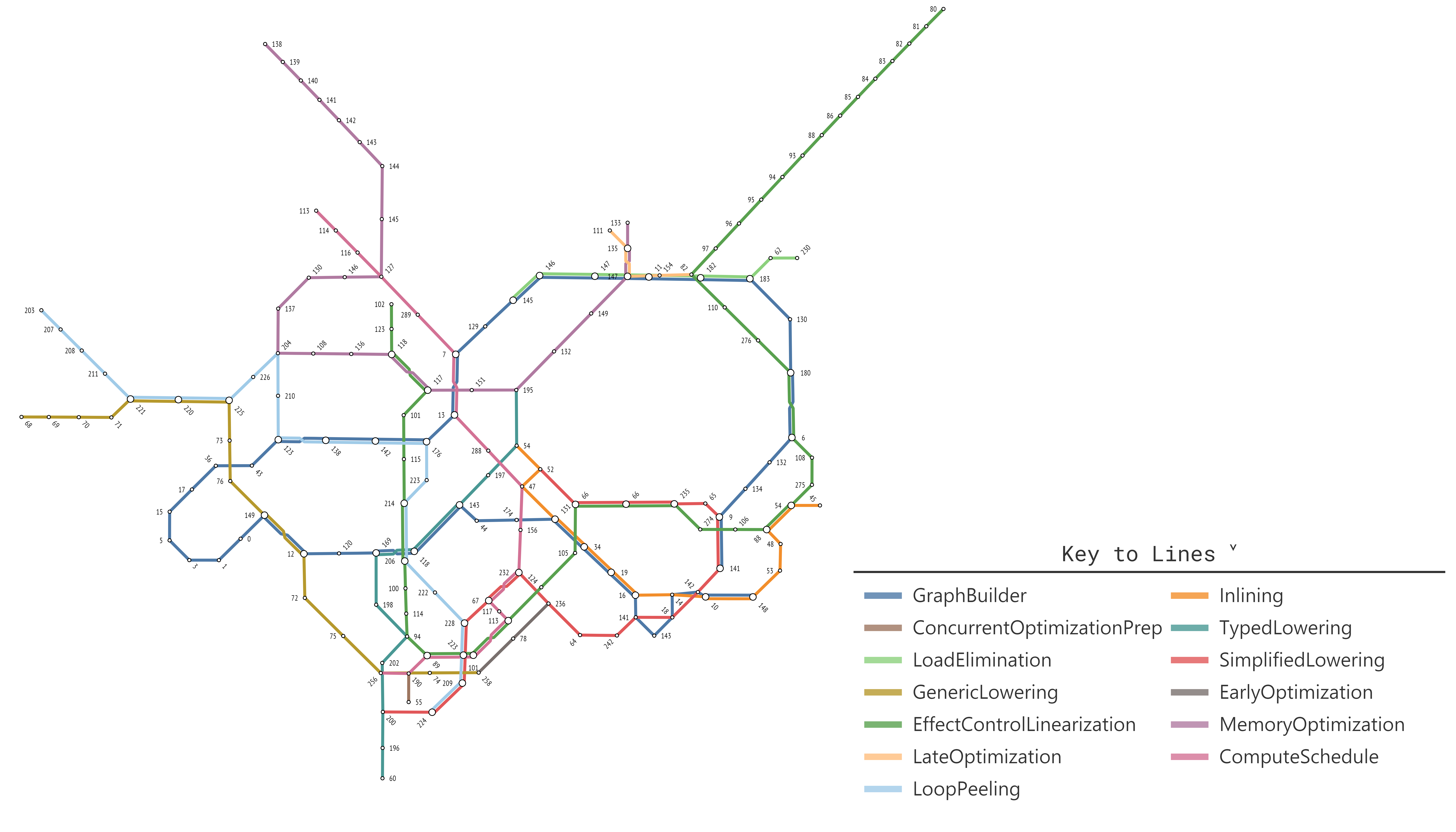}
    \caption{Bug Report 880207}
    \label{fig:880207}
\end{figure}

The bug report 880207 (Figure~\ref{fig:880207}) details an issue similar to the previous one: there is a bug in the SimplifiedLowering hyperedge where it incorrectly checks the type of "Math.expm1"~\cite{issue880207}.

Again, we can find a line in which the majority of nodes come from the other IRs. That means that the buggy program is missing  the appropriate optimization, resulting in missing nodes in the IR compared to the original IR.

\section{MetroMap Features}\label{sec:mapfeatures}

\begin{figure}[h]
    \includegraphics[width=\textwidth]{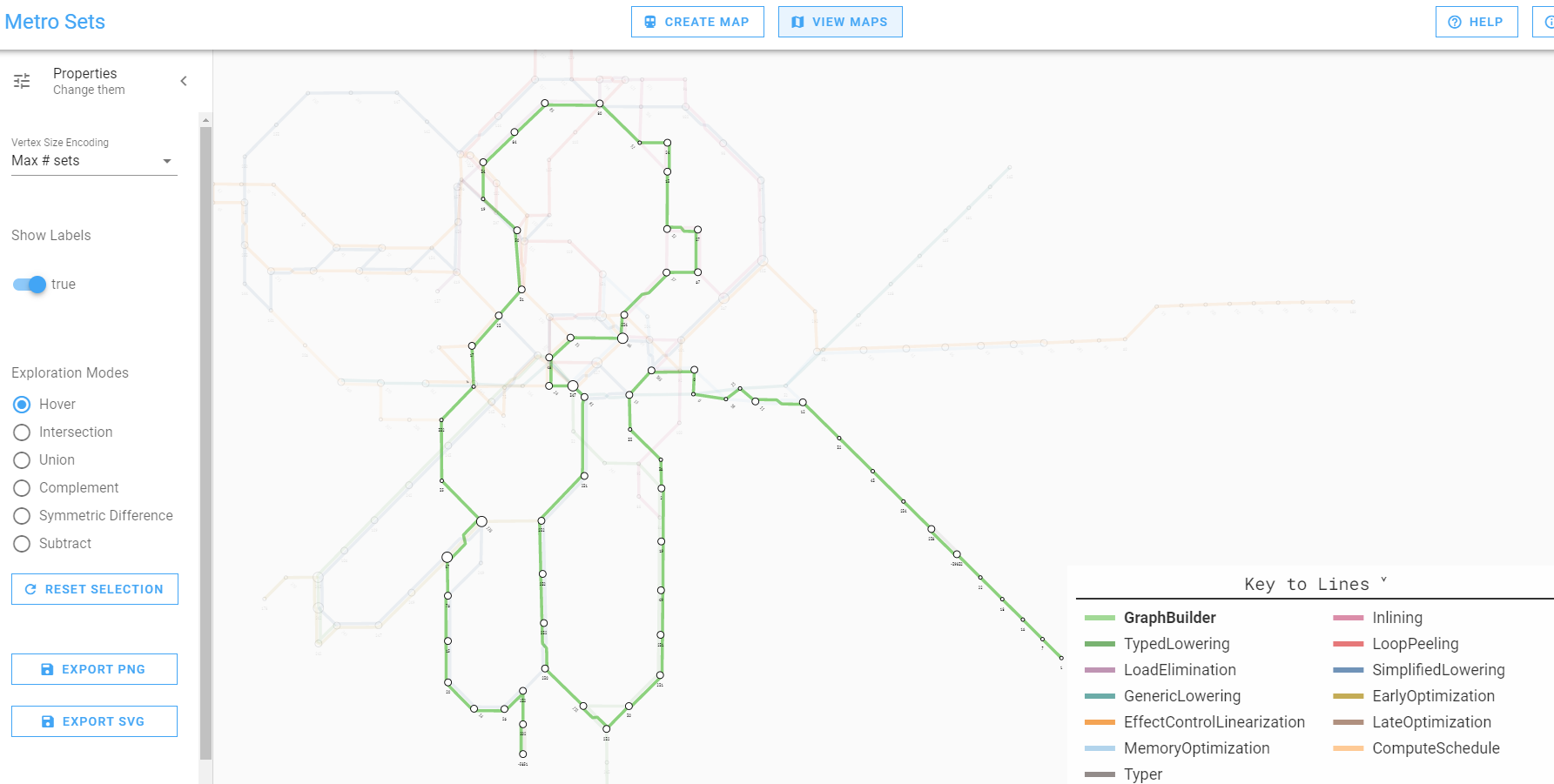}
    \caption{Hovering on a line.}
    \label{fig:linehovering}
\end{figure}

Here we briefly illustrate how different interaction with MetroSets are used to answer our list of desired questions. 

Figure~\ref{fig:linehovering} shows an example of line hovering: this results in graying out all other lines. The line name (GraphBuilder, in this example) in the ``Key to Lines" is also bolded.

\begin{figure}[hbt!]
    \includegraphics[width=.9\textwidth]{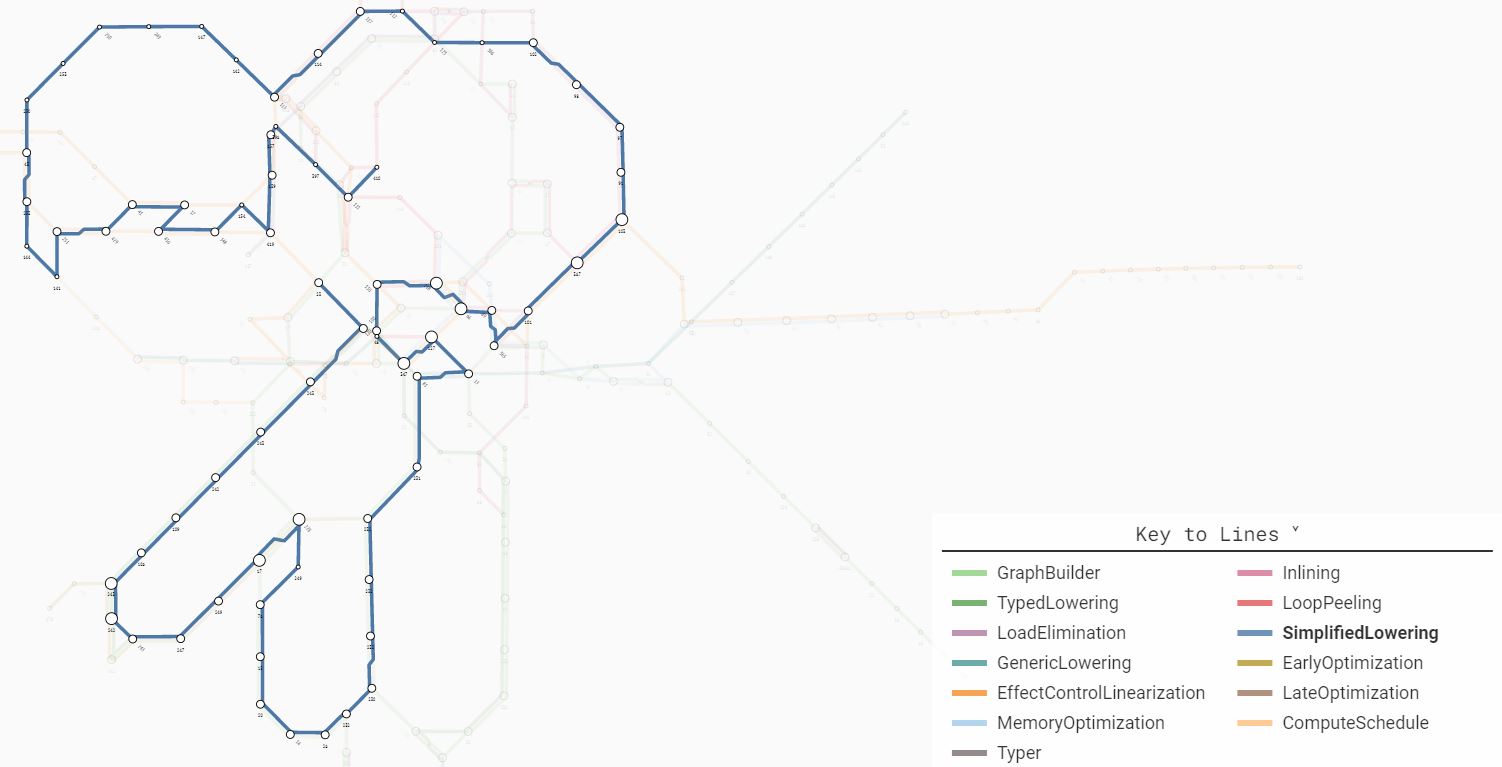}
    \caption{Most active hyperedge.}
    \label{fig:mostactivehyperedge}
\end{figure}

Figure~\ref{fig:mostactivehyperedge} shows an example of hovering over the most active hyperedge.
The example from bug report 5129 and shows that the SimplifiedLowering operation was the most active one.

\begin{figure}
    \includegraphics[width=.9\textwidth]{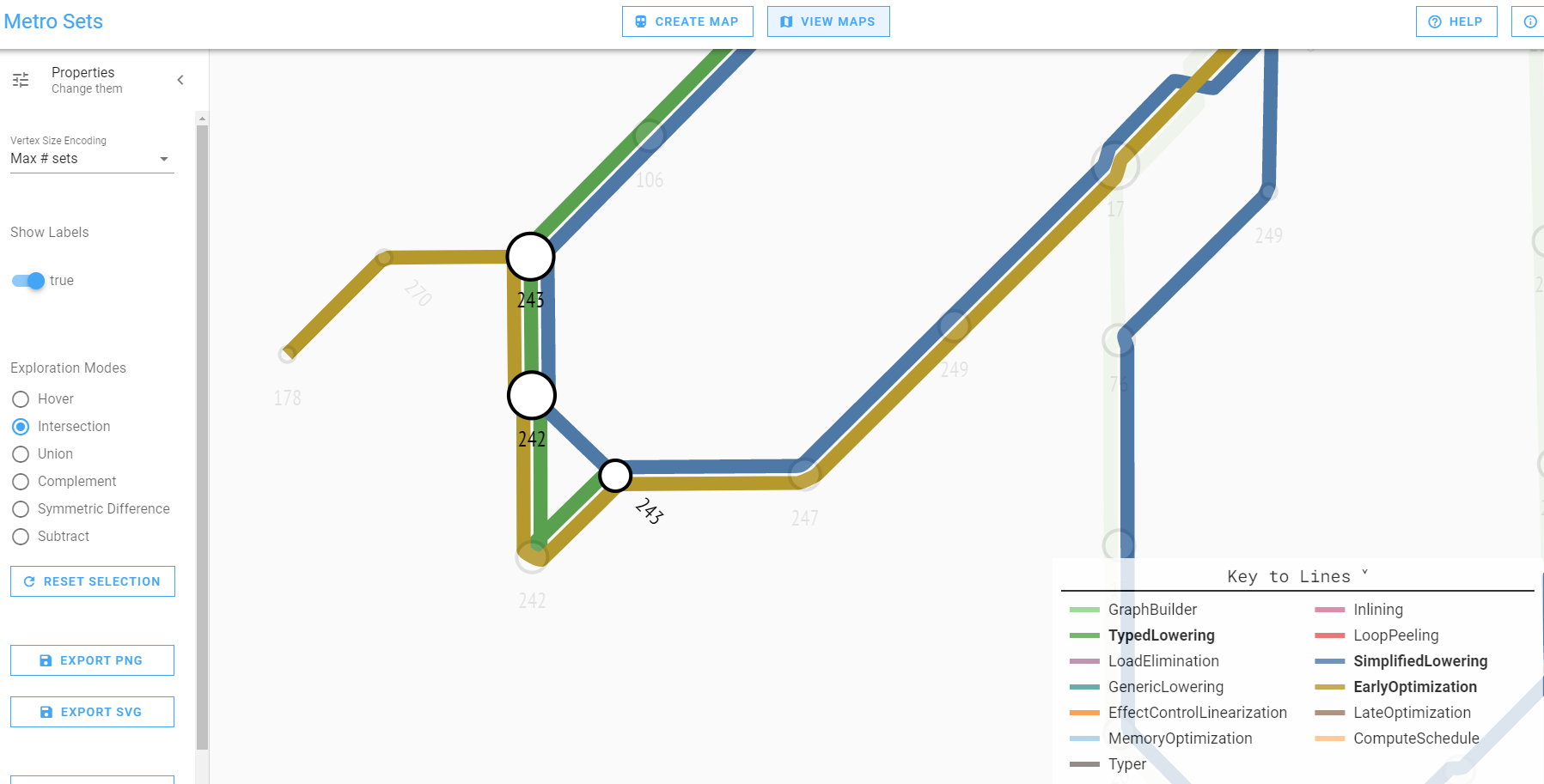}
    \caption{Intersection.}
    \label{fig:intersection}
\end{figure}

Figure~\ref{fig:intersection} shown an example of the intersection exploration mode (on the left side panel). This is particularly useful when we want to find nodes that are optimized in multiple phases. The example shows that there are 3 nodes intersecting the selected lines: TypedLowering, SimplifiedLowering, and EarlyOptimization.

\begin{figure}
    \includegraphics[width=.9\textwidth]{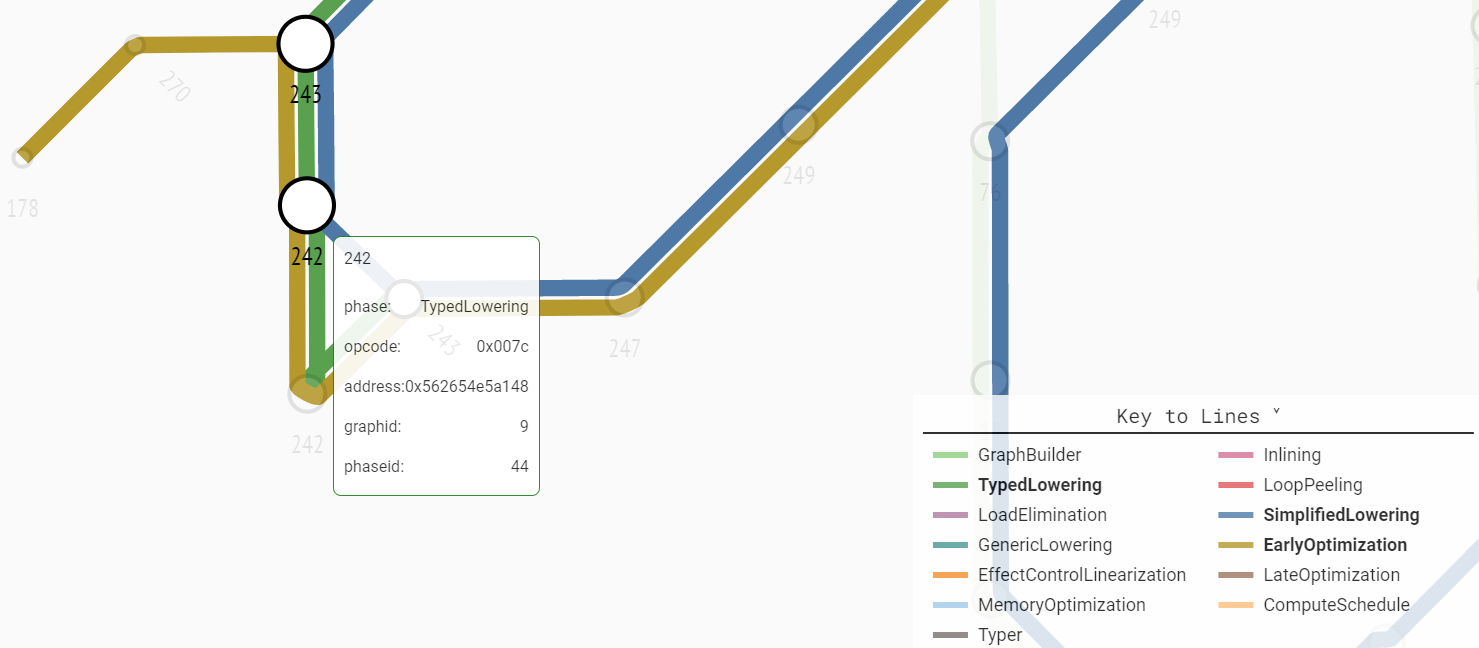}
    \caption{Node hovering}
    \label{fig:nodehover}
\end{figure}

Figure~\ref{fig:nodehover} shows an example of node hovering, which provides the attribute information associated with a node of interest. In this example, the node was generated for the arithmetic operator subtract, which can be found by  opcode $0x007c$, was generated at the TypedLowering phase and optimized at SimplifiedLowering and EarlyOptimization phases. Moreover, we can tell that this node generation and optimizations happened in late phases of the JIT compilation, identified by the phase IDs, which represent the execution order of phases.
\end{appendices}

\end{document}